\documentclass[12pt]{article}

\def\la{\mathrel{\mathpalette\fun <}}
\def\ga{\mathrel{\mathpalette\fun >}}
\def\fun#1#2{\lower3.6pt\vbox{\baselineskip0pt\lineskip.9pt
\ialign{$\mathsurround=0pt#1\hfil##\hfil$\crcr#2\crcr\sim\crcr}}}

\begin{document}
\newcommand{\beq}{\begin{equation}}
\newcommand{\eeq}{\end{equation}}
\newcommand{\Tr}{{\rm Tr}\,}
\newcommand{\nfour}{\mbox{${\cal N}\!\!=\!4\;$}}
\newcommand{\ntwo}{\mbox{${\cal N}\!\!=\!2\;$}}
\newcommand{\none}{\mbox{${\cal N}\!\!=\!1\;$}}

\renewcommand{\theequation}{\thesection.\arabic{equation}}

\begin{titlepage}
\renewcommand{\thefootnote}{\fnsymbol{footnote}}

\begin{flushright}
PNPI-TH-2411/01\\
ITEP-TH-11/01\\
TPI-MINN 01/14\\
hep-th/0103222\\

\end{flushright}

\vfil

\begin{center}
\baselineskip20pt
{\bf \Large Confinement Near Argyres-Douglas Point in \\ \ntwo QCD
and Low Energy Version \\ of AdS/CFT Correspondence}
\end{center}
\vfil

\begin{center}
{\large   Alexei Yung}

\vspace{0.3cm}

{\it Petersburg Nuclear Physics Institute, Gatchina, St. Petersburg
188300;\\ 
Institute of Experimental and Theoretical Physics, Moscow, 117259;\\
Theoretical Physics Institute, University of Minnesota, Minneapolis,
MN 55455}

\vfil

{\large\bf Abstract} \vspace*{.25cm}
\end{center}

We study Abrikosov-Nielsen-Olesen (ANO) flux tubes on the Higgs branch
of \ntwo QCD with $SU(2)$ gauge group and $N_f=2$ flavors
 of fundamental matter. In particular, we consider this theory
near Argyres-Douglas ( AD) point where the mass of monopoles
connected by these ANO strings become small.
 In this regime the effective  QED which describes the 
theory on the Higgs branch becomes strongly coupled. We argue that the 
appropriate description of the theory is in terms of long and thin
flux tubes (strings) with small tension. We interpret this
 as another example of duality
between field theory in strong coupling and string theory in weak
coupling. Then we consider the non-critical string theory for these
 flux tubes which includes fifth (Liouville) dimension. We identify
CFT at the AD point as UV fix point corresponding to
AdS metric on the 5d ``gravity'' side. The perturbation associated with
the monopole  mass term creates a kink separating UV and IR behavior.
 We estimate the renormalized string tension and show that it is
 determined by the small monopole mass. In particular, it goes to
 zero at the AD point.

\vfil

%\begin{flushleft}
%April 2001
%\end{flushleft}
\end{titlepage}

\newpage

\section{Introduction}

One of the most important physical outcomes of the Seiberg-Witten
theory \cite{SW1,SW2} is the demonstration of confinement of electric
charges via monopole condensation. Although the idea
 of confinement as a dual 
Meissner effect was suggested many years ago by Mandelstam and 't Hooft
\cite{MH} only relatively recent progress in the understanding of 
the  electomagnetic
duality in \ntwo supersymmetry allowed Seiberg and Witten  to present the
  quantitative description of this phenomenon \cite{SW1,SW2} . 

Let us recall the basic idea of Mandelstam and 't Hooft \cite{MH}.
Once monopoles (charges) condense the electric (magnetic) flux
is confined in the Abrikosov-Nielsen-Olesen (ANO) flux tube \cite{ANO}
connecting heavy trial electric (magnetic) charge
and anti-charge.
 The energy of the ANO string increases with its length. This ensures
increasing confinement potential between heavy  electric (magnetic) charge
and anti-charge.

In the Seiberg-Witten theory this confinement scenario is realized in two
possible ways. First, in the pure gauge theory near the monopole (dyon)
singularity upon breaking  \ntwo supersymmetry by the small
mass term of the adjoint matter \cite{SW1}. In this case monopoles condense
and electric charges are confined by  electric flux tubes.

Second, in the unbroken \ntwo theory with $N_f$ flavors
of the fundamental matter with degenerative masses on
 Higgs branches \cite{SW2}. In this case, say  at large masses
 of the fundamental matter electric charges condense on the Higgs branch
while magnetically charged dyons confined by  magnetic  flux tubes
\cite{HSZ,Y}.

Given  this progress it is still unclear if the confinement in the 
Seiberg-Witten theory can be taken as a  model (at least
qualitatively) for the confinement in QCD. 
The point is that the confinement
in the Seiberg-Witten theory has several unwanted properties we do
not expect to have in QCD (see \cite{YR} for a recent review).

In particular, one group of such properties of confinement
 in Seiberg-Witten theory  is related to the large 
value of the ANO string tension $T$. In terms of the
 photon mass $m_{\gamma}$ of the effective 
low energy \ntwo QED it is given by
\beq
T\sim \frac{m_{\gamma}^2}{g^2}.
\label{tint}
\eeq
At small values of QED coupling $g^2$ it is much larger then $m_{\gamma}$.
As a consequence the low energy hadron spectrum, say near the monopole
point consists of relatively light photon and monopoles 
(which we would like to interpret as glueballs) of mass $m_{\gamma}$
and heavy hadrons built of quarks connected by ANO strings with mass
of order of $\sqrt{T}$ \cite{YR}\footnote{This is true if we take
the bare mass of quarks to be small, the possibility discussed in
detail below.}.
 In contrast, in QCD we have light
$\bar{q}q$ states while the candidates for glueballs are relatively
heavier.

Another problem related to the large value 
of ANO string tension (\ref{tint})
is the non-linear behavior of Regge trajectories. The transverse
size of a string is of order of $1/m_{\gamma}$. This is much larger then
its length which is of order of $1/\sqrt{T}$ for small hadron spins $j$.
 Therefore, the string is not
developed (grows thick and short) and the  $\bar{q}q$ state looks more like
spherically symmetric soliton rather than a string. This is the reason for
the non-linear behavior of Regge trajectories in wide region of spins
$j\la 1/g^2$ \cite{YR}.

The purpose of this paper is to overcome the 
above mentioned problems 
related to the large value of the string tension (\ref{tint}).
We suggest a regime in the Seiberg-Witten theory
in which ANO strings becomes light (almost tensionless).

To be more specific, we consider the second  of the above mentioned
scenarios of confinement which arises on the Higgs branch. Namely,
we consider \ntwo gauge theory with $SU(2)$ gauge group and
two hypermultiplets of the fundamental matter (we call them quarks).
If the masses of quarks are equal the Higgs branch is developed.
It touches the Coulomb branch at the singular point where some
quarks become massless \cite{SW2} 
(we give a brief review of Higgs branches
in the next section). The effective low energy description
of the theory near the root of Higgs branch is given by \ntwo QED.
When scalar quarks develop vev's on the Higgs branch the effective
QED is in the Higgs phase. The formation of ANO flux tubes
in this vacuum leads to confinement of monopoles (any dyons with
non-zero magnetic charge) \cite{HSZ,Y}
\footnote{Strictly speaking this 
terminology refers to large values of bare quark  masses. For small
quark masses (below Argyres-Douglas point)
quantum numbers of particles are changed because of monodromies \cite{SW1}
and now monopoles are condensed at the Higgs branch while 
electric charges are confined \cite{BF}.  To avoid
confusion in this paper we use the terminology which refers to
 large quark masses. }
 
At large values of bare quark mass $m$ monopoles are heavy and can
be viewed as heavy trial particles to probe confinement. 
It is a challenging
problem to see what happens to confinement if we reduce the monopole
mass and  make monopoles   dynamical. If ANO flux tubes still exist
in this regime and have finite transverse size we would still expect
confinement of monopoles to occur, although the Wilson loop does not
show the area law any longer (flux tubes can be broken by light
monopole-anti-monopole pairs). In this setup the problem of confinement
becomes similar to that in QCD.

In this paper we study what happen to ANO flux 
tubes on the Higgs branch of
the Seiberg-Witten theory once we reduce $m$ and eventually come
close to the Argyres-Douglas (AD) point.
 These points were originally introduced in the moduli/parameter
space of \ntwo theories as 
points  where two singularities with mutually non-local light states
on the Coulomb branch
coalesce~\cite{AD,APSW,hori}. It is believed that the theory in
 the AD point flows in the infrared to a nontrivial
superconformal theory (let us call it $CFT_{AD}$).
  We consider the theory on the Higgs branch
 where  massless scalar quarks develop vev $v$.
 AD point corresponds
to the value of the bare quark mass $m$ equal to some
critical value $m_{AD}$ at which monopoles also become  massless.

When we come close to the AD point our effective QED description
is no longer valid because QED enters a strong coupling regime.
Our main proposal in this paper is that the ANO string tension
$T$ becomes small, $T\ll v^2$. We give  arguments based on
what we know about the effective sigma model on the Higgs branch  in 
favor of this conjecture.
 As we explained above the small value of 
string tension  eliminates certain important ``disadvantages''
of confinement in the Seiberg-Witten theory making it similar to
the one we expect in QCD.

Once $\sqrt{T}$ is much smaller then the inverse
 transverse size of the flux tube
the correct low energy description of the theory is in terms
of long and thin strings. We apply methods of non-critical string
theory developed mostly by Polyakov \cite{P1,P} to our ANO flux tube.
We consider this as another example of duality between field theory
(QED, in the case at hand)
 at strong coupling and string theory at weak coupling. 

The non-critical string theory contains curved fifth (Liouville)
 coordinate $u$ \cite{P1,P}
which is associated with RG scale in field theory \cite{M}. We suggest
a low energy version of AdS/CFT correspondence \cite{M,GKP,W} in which
the AdS metric  at large $u$ on the  5d ``gravity'' side
 corresponds to $ CFT_{AD}$ 
on the field theory side in the UV limit.
Note, that we use the word gravity in quotation marks here
 because it has nothing to do
with the real gravity at the Planck scale.
 In this paper we discuss ``gravity
inside hadrons'' \cite{P} which appears at the hadron scale $\sim \sqrt{T}$.

Then we consider breaking of the conformal invariance induced by
small monopole mass  near the AD point (it is determined by $m-m_{AD}$).
On the ``gravity'' side we associate this perturbation with a scalar field
in 5d ``gravity''. We study 5d ``gravity'' equations of motion 
with this scalar included and
find the scale $u_{CB}$ at which the kink separating
 the UV and IR behavior destroys the AdS metric. Using $u_{CB}$ we estimate
the renormalized  ANO string tension. It turns out to be small, proportional
to the  monopole mass. This result shows selfconsistency
of  our initial conjecture.
In particular, $T$ goes to zero at the AD point.

Then we verify that  the radius of AdS space is large in string units
justifying the ``gravity'' approximation .\footnote{Note, we do not
use the logic of large $N$ \cite{M} in this paper.} We also make  two
estimates of the string coupling constant $g_s$, one from the ``gravity''
side and another
one from the field theory side. Although our accuracy is not
enough to show quantitative agreement both calculations shows that 
 $g_s $   is small, $g_s\ll 1$.

The paper is organized as follows. In sect. 2 we review quasiclassical
results on ANO flux tubes on  Higgs branches of Seiberg-Witten theory.
In sect. 3 we discuss field/string theory duality near the AD point and
introduce 5d ``gravity'' description. In sect. 4 we consider perturbation
of $AdS_5$ metric and estimate the renormalized string tension.
Sect. 5 contains our conclusions and discussion.

\section{ANO strings on Higgs branch in the quasiclassical regime}
\setcounter{equation}{0}

In this section we review quasiclassical results obtained for ANO
flux tubes on Higgs branch of \ntwo QCD with gauge group $SU(2)$ and
$N_f=2$ flavors of fundamental matter (quarks) with common bare  mass
parameter $m$ \cite{Y}. First, we briefly review the effective theory
on the Higgs branch \cite{SW2}.

\subsection{Higgs branch}

The \ntwo vector multiplet of the  theory at hand
on the component level consists of  the
gauge field $A^a_\mu$, two Weyl fermions $\lambda^{\alpha a}_1$
and $\lambda^{\alpha a}_2$ $(\alpha=1,2)$ and the complex scalar
$\varphi^a$, where $a=1,2,3$ is the color index. Fermions form
a doublet  $\lambda^{\alpha a}_f$ with respect to global $SU(2)_R$
group, $f=1,2$.

The scalar potential of this theory has a flat direction. 
The adjoint scalar field  develop an arbitrary vev along this
direction breaking $SU(2)$ gauge group down to $U(1)$. We choose
$\langle\varphi^a\rangle=\delta^{a3}\langle a\rangle$. The
complex parameter $\langle a\rangle$ parameterize the 
Coulomb branch. The low energy effective
theory  generically contains only the photon $A_\mu=A^3_\mu$ and its
superpartners: two Weyl fermions $\lambda^{3\alpha}_f$
and the complex scalar $a$. This is massless short vector \ntwo
multiplet. It contains 4 boson + 4 fermion states.
  W-boson and its superparthners are massive with masses
 of order of $\langle a\rangle$.

Quark hypermultiplets has the following structure. They
consist of complex scalars $q^{kfA}$ and fermions $\psi^{k\alpha A}$,
$\tilde \psi^{\dot{\alpha}}_{Ak}$, where  $k=1,2$ 
is the color index and $A=1,\ldots,N_F$ is the flavor one.
Scalars form a doublet with respect to $SU(2)_R$ group.
All these states are in the BPS short representations of
\ntwo algebra  on the Coulomb
branch with $4\times N_c\times  N_f=16$ real boson states (+ 16
fermion states).

Coulomb branch has three singular points where
monopoles , dyons or charges become massless. Two
of them correspond to monopole and dyon singularities of the
pure gauge theory. Their positions on the Coulomb branch are
given by \cite{SW2}
\begin{equation}
\label{mds}
u_{m,d}\ =\ \pm\ 2m\Lambda_2-\frac12\Lambda^2_2\ ,
\end{equation}
where $u=\frac12\langle\varphi^{a^2}\rangle$ and $\Lambda_2$ is
the scale of the theory with $N_f=2$.
In the large $m$ limit $u_{m,d}$ are
approximately given by their values in the pure gauge theory
$u_{m,d}\simeq\pm2m\Lambda_2=\pm2\Lambda^2$, where $\Lambda$
is the scale of $N_f=0$ theory.

The charge singularity corresponds to the point where half of quark
states
becomes massless. We denote them $q^{fA}$ and  $\psi^{\alpha A}$,
$\tilde \psi^{\dot{\alpha}}_{A}$ dropping the color index. They form
$N_f=2$ short hypermultiplets with $4\times  N_f=8$ real boson states.
The rest of  quark states acquire large mass $2m$ and we ignore them
in the low energy description. 
The charge singularity appears  at the point
\begin{equation}
\label{acs}
a\ =\ -\ \sqrt2\ m\ 
\end{equation}
on the Coulomb branch.
 In terms of variable $u$ (\ref{acs}) reads
\begin{equation}
\label{cs}
u_c\ =\ m^2+\frac12\Lambda_2^2.
\eeq
Strictly speaking, we have $2+N_f=4$ singularities on the
Coulomb branch. However, two of them  coincides for the case of
two flavors of matter with the same mass.

The effective theory on the Coulomb branch near the charge
singularity (\ref{acs}) is given by 
${\cal N}=2$ QED with light matter fields
$q^{fA}$, and their superparthners  as well as the
photon multiplet.

The charge singularity (\ref{acs}) is the root of the Higgs
branch \cite{SW2}. To find it we impose 
 $D$-term
and $F$-term conditions which look like
\begin{equation}
\label{df}
\bar q_{Ap}(\tau^m)^p_f\ q^{fA}\ =\ 0, \quad m=1,2,3.
\end{equation}
(Here $m$ is an adjoint $SU(2)_R$ index, not to be
confused with color indices.)
This equation  determines the Higgs
branch (manifold with $\langle q\rangle\neq0$) which touches the
Coulomb branch at the point (\ref{acs}).
It has non-trivial solutions for $N_f\ge 2$ \cite{SW2}. This is 
the reason why we choose $N_f= 2$ for our discussion.

The low energy theory for boson fields near the root of the Higgs
branch is given by
\begin{equation}
\label{rtact}
S^{\rm root}_{\rm boson}=\int d^4x\left\{\frac1{4g^2}
F^2_{\mu\nu}+\bar\nabla_\mu\bar q_{Af}\nabla_\mu q^{fA}+
\frac{g^2}8[\mbox{ Tr }(\bar q\tau^m q)]^2\right\},
\end{equation}
where trace is calculated over flavor and $SU(2)_R$ indices.
Here $\nabla_\mu=\partial_\mu-in_eA_\mu$, $\bar\nabla_\mu=
\partial_\mu+in_eA_\mu$, the electric charge $n_e=1/2$ for
fundamental matter fields.

This is an Abelian Higgs model with last interaction term coming
from the elimination of $D$ and $F$ terms. The QED coupling
constant $g^2$ is small near the root of the Higgs branch if
$m$ is not close to its AD value $m_{AD}$.
 The effective theory (\ref{rtact}) is
correct on the Coulomb branch near the root of the Higgs branch
(\ref{acs}) or on the Higgs branch not far away from the origin
at $\langle q\rangle=0$.

Once $|\langle q\rangle|^2 = v^2 \ne 0$ on the Higgs
 branch the $U(1)$ gauge
group in (\ref{rtact}) is broken and the photon acquires the mass
\beq
\label{mg}
m^2_{\gamma}=\frac{1}{2}g^2 v^2
\eeq

It is clear that the last term in (\ref{rtact}) is zero on  fields
$q$ which satisfy constraint (\ref{df}). This means that moduli
fields which develop vev's on the Higgs branch are massless, as
expected.
 It turns out that there are four
real moduli fields $q$ (out of 8) which satisfy the constraint
(\ref{df}) \cite{SW2}. They correspond to the lowest components of  one
 short hypermultiplet.

 The other quark fields (4 real boson
 states + fermions) acquire  the  mass of  the photon (\ref{mg}).
Together with states from the photon multiplet they form
one long (non-BPS) \ntwo multiplet (cf. \cite{VY}). It has 8 boson
+ 8 fermion states. The reason why the long multiplet appears is 
easy to understand. Electric charge is screened by the quark
condensation on the Higgs branch. Therefore, the central
 charges of \ntwo algebra
are zero now. That means that there is no
BPS particles any longer. Note, that the multiplet of moduli
fields is short (4 boson + 4 fermion states) because it is 
massless.

We can parameterize  massless moduli fields  as
\begin{equation}
\label{mod}
q^{f\dot A}(x)\ =\ \frac1{\sqrt2}\ \sigma^{f\dot A}_\alpha
\phi_\alpha(x) e^{i\alpha(x)}\ .
\end{equation}
Here $\phi_\alpha(x), \alpha=1\ldots4$ are four real moduli
fields. It is clear that fields (\ref{mod}) solve (\ref{df}). The common
phase $\alpha(x)$ in (\ref{mod}) is the U(1) gauge phase. Once
$\langle\phi_\alpha\rangle=v_\alpha\neq0$ on the Higgs branch the U(1)
group is broken and $\alpha(x)$ is eaten by the Higgs  mechanism.
Say, in the unitary gauge $\alpha(x)=0$. In the next subsection
we consider vortex solution for the model (\ref{rtact}). Then
$\alpha(x)$ is determined by the behavior of the gauge field at
the infinity.

Once $v_\alpha\neq0$ we expect monopoles (they are heavy at
$m\gg\Lambda_2$) to confine via formation of vortices which carry
the magnetic flux. The peculiar feature of the theory (\ref{rtact}) is
the absence of the Higgs potential for the
moduli fields $\phi_{\alpha}$. Therefore, the Higgs phase of
of the theory in
(\ref{rtact}) is the limiting case of type I superconductor with the
vanishing Higgs mass. In the next subsection we will review the
peculiar features of ANO vortices in this model.

If we consider the low energy limit of the theory at 
energies much less then the photon mass (\ref{mg})
  we
can integrate out massive fields. Then the effective theory is a
$\sigma$-model for massless fields $\phi_\alpha$ which belong to
4-dimensional hyper--Kahler manifold  $R^4/Z_2$. The metric of
this $\sigma$-model is flat \cite{SW2,APS}, there are, however,
higher derivative corrections induced by instantons \cite{Y2,Y3}.

\subsection{ANO string}

Now let us  review classical solution for 
 ANO vortices in the model (\ref{rtact}) \cite{Y}.
Without loss of generality we take vev's of $\phi_\alpha$
$v_\alpha=(v,0,0,0)$. Moreover, we drop fields $\phi_2,\phi_3$ and
$\phi_4$ together with massive scalars 
 from (\ref{rtact}) because they are irrelevant for the purpose of
finding classical vortex solutions. Thus, we arrive at the
standard Abelian Higgs model
\begin{equation}
\label{ah}
S_{AH}=\int d^4x\left\{\frac1{4g^2}\,F^2_{\mu\nu}+|\nabla_\mu
q|^2+\lambda(|q|^2-v^2)^2\right\},
\end{equation}
for the single complex field $q$  with  quartic coupling $\lambda=0$.
Here
\begin{equation}
q(x)\ =\ \phi_1(x)\ e^{i\alpha(x)}\ .
\end{equation}
Following \cite{Y} consider first the model (\ref{ah}) with small
$\lambda$, so that the Higgs mass $m_H\ll m_\gamma$
  ($m_H^2=4\lambda v^2$).
 Then  we take the limit $m_H\to0$.

To the leading order in $\log m_\gamma/m_H$ the vortex
solution has the following structure in
the plane orthogonal to the string axis \cite{Y}. The
electromagnetic field is confined in a core with the radius
\begin{equation}
\label{rad}
R^2_g\ \sim\ \frac1{m^2_\gamma}\ln^2\frac{m_\gamma}{m_H}\ .
\end{equation}
The scalar field is close to zero inside the core. Instead,
outside the core, the electromagnetic field is vanishingly
small while the scalar field slowly 
(logarithmically) approaches its boundary
value $v$.
 The result for the string tension is \cite{Y} 
\begin{equation}
\label{rten}
T_{\lambda}\ =\ \frac{2\pi v^2}{\ln m_\gamma/m_H}\ .
\end{equation}
The main contribution to the tension in (\ref{rten})
 comes from the logarithmic ``tail'' of the scalar field.

The results in (\ref{rad}), (\ref{rten}) mean that if we naively take the
limit $m_H\to0$ the string becomes infinitely thick and its
tension goes to zero \cite{Y}. This means that there are no
strings in the limit $m_H=0$.  The absence of ANO
strings in theories with flat Higgs potential was first noticed in
\cite{R,ARH}.

One might think that the 
 absence of ANO strings means that there is no
  confinement on Higgs branches. As
we will see now this is not  the case \cite{Y}.
 So far we have considered infinitely long ANO strings. However
the setup for the confinement problem is slightly different
\cite{Y}. We have to consider monopole--anti-monopole pair at
large but finite separation $L$. Our aim is to take the limit
$m_H\to0$. To do so let us consider ANO string of the finite
length $L$ within the region
\begin{equation}
\frac1{m_\gamma}\ \ll\ L\ \ll\ \frac1{m_H}\ .
\end{equation}
Then it turns out that $1/L$ plays the role of the $IR$-cutoff
 in Eqs. (\ref{rad}) and (\ref{rten}) instead of $m_H$ \cite{Y}.
Now we can safely put $m_H=0$.

The result for the electromagnetic core of the vortex becomes
\begin{equation}
\label{ts}
R^2_g\ \sim\ \frac1{m^2_\gamma}\ln^2m_\gamma L\ ,
\end{equation}
while its string tension is given by \cite{Y}
\begin{equation}
\label{ct}
T_0\ =\ \frac{2\pi v^2}{\ln m_\gamma L}\ .
\end{equation}

We see that the ANO string becomes "thick" but still its
transverse size $R_g$ is much less than its length $L$, $R_g\ll
L$. As a result the potential between heavy well separated
monopole and anti-monopole is still  confining but is
no longer linear in $L$. It behaves
as \cite{Y}
\begin{equation}
V(L)\ =\ 2\pi v^2\, \frac L{\ln m_\gamma L}\ .
\end{equation}
 As soon as the
potential $V(L)$ is an order parameter which distinguishes 
different phases of a theory
(see, for example, review
\cite{IS}) we conclude that we have a new confining phase on the
Higgs branch of the Seiberg--Witten theory. It is clear that
this phase can arise only in supersymmetric theories
because we do not have Higgs branches without supersymmetry.

These quasiclassical results are valid if the effective QED coupling is 
small, $g^2\ll 1$. It is small if two conditions are satisfied.
First, $v\ll \Lambda_2$ ensures that W-bosons are massive
and we can ignore them and use the effective QED description.
Second, is that $m$ is not close to $m_{AD}$ , which ensures 
that monopoles/dyons are massive. In the next section we relax
the second condition and study what happen to ANO strings if
we go close to AD point.

Now let us comment on why we choose this relatively complicated 
type of ANO string for our study of what happens to confinement near
AD point. The reason is that \ntwo supersymmetry is unbroken 
on the Higgs branch. We will use this heavily in the next section.
Alternatively, one could consider  the confinement scenario
near the monopole point which arises upon breaking  \ntwo
supersymmetry  down to \none by the mass term of the adjoint matter.
In this case ANO strings appear to be ``almost'' BPS saturated
once the breaking is small \cite{HSZ,VY}. However, if we 
 reduce quark mass going to AD point the monopole condensate vanishes
showing deconfinement \cite{GVY}. We can try to find a regime in which
monopole condensate is fixed while quark mass goes to zero. However,
 it is easy to show that in this regime  \ntwo breaking becomes strong
\cite{YR}.

\section{ANO string at the AD point}
\setcounter{equation}{0}

In this section we discuss what happen to ANO strings if we go
close to AD point. On the Coulomb branch the mass of the
 monopole  is given by the BPS formula \cite{SW1}
 $m_m=\sqrt{2}|a_D|$, where $a_D$ is the dual 
variable to $a$. Now let us take $m$ close to $m_{AD}$ so the  monopole
singularity collides with the root of the  Higgs branch
 (charge singularity).
From (\ref{mds}) and (\ref{cs}) we learn that
\beq
m_{AD}=\Lambda_2.
\eeq
Since $a_D=0$ in the monopole point, while charges
are massless in the charge singularity  we have simultaneously
both charges and monopoles becoming massless at $m=\Lambda_2$.
The theory at AD point on the Coulomb branch flows in the 
IR to a non-trivial interacting fixed point \cite{AD,APSW} which we call
$CFT_{AD}^C$. The superscript $C$ indicates here that we are
 talking about the 
CFT on the Coulomb branch. The conformal dimension of $a_D$ equals to 1
near AD point, while the conformal dimension of $(m-\Lambda_2)$ is
2/3 \cite{APSW}. Thus, we conclude that the monopole mass behaves
as
\beq
\label{mm}
 m_m\sim \frac{(m-\Lambda_2)^{3/2}}{\Lambda_2^{1/2}}
\eeq
when $m\to \Lambda_2$.

 Let us go to the Higgs branch taking $v\ne 0$. Now  we have two
scales in the problem $v$ and $m_m$ determined by
 $m-\Lambda_2$ via (\ref{mm}). If $m_m \gg v$
then monopoles are heavy. The monopole multiplet contains 4 boson
+4 fermion states. Still monopoles cannot be BPS saturated 
on the Higgs branch because central charges  of \ntwo algebra are zero.
They are confined by ANO flux tubes and instead of monopoles we
see hadrons built of open ANO
 string states with monopoles and anti-monopoles
attached to  string ends. These are, of course, non-BPS states.
Note, that as we discussed in the Introduction, in fact, strings are  not
developed in this regime. They grow short and thick  
provided  the effective QED coupling is small, see (\ref{ts}), (\ref{ct}).

More specifically, if $m_m \gg v$
the effective coupling is of order of
\beq
\label{chb}
g^2\sim -\frac1{\log (m_{\gamma}/\Lambda_2)},
\eeq
frozen at the photon mass scale. Thus, $g^2\ll 1$ at $v\ll \Lambda_2$.
Instead, near the monopole point on the Coulomb branch the 
dual coupling constant is small, thus $g^2$ is large
\beq
\label{cm}
g^2\sim -\log (m_m/\Lambda_2).
\eeq
It is clear from last two equations that the QED coupling constant on the 
Higgs branch increases as we reduce $m-\Lambda_2$ and eventually we enter
the strong coupling regime, $g^2\sim 1$ at $m_m \sim v$.

\subsection{Tensionless strings and QED/string theory duality}

Now let us discuss what happens if we reduce $m_m$ well below $v$ making
monopole much lighter then photon. At first glance, the natural guess is
that ANO string tension stays large, 
\beq
\label{nt}
T\sim \frac{v^2}{\log vL}
\eeq
as it is suggested by the quasiclassical result (\ref{ct}). This seems
 natural because the string is ``built'' of quarks and electromagnetic
field and seems to have nothing to do with monopoles. However,
 as we will now show the guess in (\ref{nt}) is not correct.

Suppose we keep $m_m\ll v$ but do not
go directly to AD point and  integrate out hadrons built of monopoles
together with photon multiplet. Then  we are left with a sigma model for
massless quark moduli.  As soon as the Higgs branch
is a hyper-Kahler manifold its metric is determined uniquely and cannot
receive corrections. In fact, it is known to be flat \cite{SW2,APS}. 
However, there are higher derivative corrections.  It is clear that
higher derivative corrections encode all the information about
massive states we have integrated out.

On  dimensional grounds higher derivative corrections can go
in powers of
\beq
\label{vcor}
\frac{\partial^2}{v^2} 
\eeq
or in powers of 
\beq
\label{mcor}
\frac{\partial^2}{(m-\Lambda_2)^2}.
\eeq 
The difference  $(m-\Lambda_2)$ appears here because it is clear that
the theory can have singularities only at the AD value of $m$, 
$m=\Lambda_2$. At any other values of $m$ the theory is smooth.

Higher derivative corrections of type (\ref{vcor}) become singular
at $v=0$ (on  the Coulomb branch) showing that
certain states become massless in this limit. Photon multiplet
is an example of such a state.
Higher derivative corrections of type (\ref{mcor}), if present,
would signal that some states become massless at AD point
on the Higgs branch at non-zero $v$. If (\ref{nt}) were correct
there would be no such corrections because  all hadrons built
 of monopoles are heavy with masses
of order of $\sqrt{T}\sim v$ and cannot produce 
singularities of type (\ref{mcor}).

Now the question is whether there are higher derivative corrections
of type (\ref{mcor}). This problem was studied in \cite{Y2}.
In particular, in \cite{Y2} higher derivative corrections on the
Higgs branch induced by one instanton were calculated.
Consider  large $m$, $m\gg \Lambda_2$ far away from the 
AD point. Then the Higgs branch is in the weak coupling and
quasi-classical methods can be applied. The holomorphic one instanton
induced corrections
 appear to be
non-zero and proportional to powers of $\partial /m$ and $\Lambda_2/m$.
Note, that $v$ is considered large in \cite{Y2}, $v\gg m$. Thus, these
instanton corrections are really of type (\ref{mcor}) rather then
of type (\ref{vcor}).
Now if we reduce $m$ going to AD point these corrections blow up
showing a singularity at $m=\Lambda_2$.
 This singularity should correspond to some
extra states (besides quark moduli) becoming 
massless. These extra states
cannot be just monopoles because monopoles are in the confinement
phase. They  are bound into hadrons by  ANO strings.

 The plausible suggestion
is that some hadrons built of monopoles  (ANO string
states) become massless. This could happen only if the ANO string becomes 
tensionless at the AD point. Thus, we conclude that
  (\ref{nt}) is not correct and suggest instead that
\beq
\label{smallt}
T\ll v^2
\eeq
at $m_m\ll v$. In particularly, we need
\beq
\label{dec}
T( m_m\to 0) \to 0.
\eeq

Note, that alternatively, we could suggest that the string tension stays
large but some of string states become massless. Definitely this would
be the case for a critical string. However, for a non-critical string
(moreover, for a string without world sheet conformal invariance)
this hardly can happen.
 
In particularly, (\ref{dec}) means that strictly at the  AD point
the theory flows in the IR to 
 a non-trivial conformal field theory of interacting massless
quark moduli with massless string states. We call this theory
  $CFT_{AD}^H$ where $H$ stands for the Higgs branch. This is a
descendant of $CFT_{AD}^C$ which is a CFT at the AD point on the 
Coulomb branch.

The conclusion in eqs. (\ref{smallt}), (\ref{dec}) is quite a dramatic one.
It means that we loose confinement as we reduce  mass of confining matter.
In particular, (\ref{dec}) means that size of hadrons built of
monopoles becomes infinite as we approach AD point.
   This is in a contradiction with the standard point of view
that, say, confinement in QCD is insensitive to the quark mass.
We assume the proposal  (\ref{smallt}), (\ref{dec}) below 
in this paper and check its selfconsistency.

The proposed behavior  although surprising is quite
similar to what happens to monopoles in the Seiberg-Witten theory.
At certain point on the Coulomb branch the monopole becomes massless.
Its size stays small, of order of the inverse W-boson mass so we can
consider monopole as a point-like particle to be included in the effective 
low energy  theory. Similar to that,
 in the case at hand, at certain point
of the parameter space ($m=\Lambda_2$) the ANO string becomes 
tensionless, while its transverse size $R_g$ remains small, determined
by the mass of the photon ($\sim v^{-1}$).

This means that ANO  strings are now long and thin (with typical length
$L\sim 1/\sqrt{T}$). This is exactly what is assumed for a string
 in the  string theory.  Thus, we expect that 
appropriate description of our theory is in terms of non-critical theory
of ANO strings. We interpret this as a duality between field theory (
\ntwo QED) at strong coupling and the string theory at weak coupling
\footnote{We confirm in the next section that the string
coupling $g_s$ appears to be small.}.

In the conclusion of this subsection let us compare the physics on 
the Higgs branch at the AD point with the one we have at the AD point
of $N_f=1$ theory with \ntwo supersymmetry broken down to \none by 
the mass term of the adjoint matter \cite{GVY}.  Suppose we are 
at the monopole vacuum turning the quark mass parameter 
in a way to  ensure 
the collision of the monopole vacuum with the charge one.
Then the monopole condensate goes to zero in the AD point \cite{GVY}.
 The monopole condensate
sets the mass scale of all light states in the theory, thus, all of them 
become massless (including photon). In particular, the ANO string becomes
tensionless, however its transverse size (given by the inverse photon mass)
goes to infinity. Therefore, it cannot be considered as a stringy
object, in fact, it disappears \cite{GVY}.

 On the Higgs branch in the theory at hand the physics is quite different.
At the AD point ANO strings becomes tensionless too, but
their transverse size remains small, of order of $v^{-1}$. The string
remains a localized object in  transverse directions. In fact, the mass
of the photon ($\sim v$) plays now the role of the UV cutoff for our
low energy effective string theory.
Below
we describe this string theory and use it to calculate the string tension
$T$.

\subsection{5d string theory and AdS metric}

One can develop a string theory representation for the ANO string in the 
quasiclassical regime $g^2\ll 1$. 
 This is done  in
\cite{O,BS} and \cite{Y} for  cases of strings in the type II
superconductor, BPS-strings and strings in the type I
superconductor respectively. The common feature of these
representations is that the leading term of the world sheet
action is the Nambu--Goto term
\begin{equation}
\label{ng}
S_{\rm str}=\ T\int d^2\sigma\left\{\sqrt{ g^{ind}}+\ \mbox{ higher
derivatives }\right\},
\end{equation}
where $g^{ind}_{ij}=\partial_ix_\mu\partial_jx_\mu$ is the induced metric
$(i,j=1,2)$. Higher derivative corrections in (\ref{ng}) include the 
Jacobian term  \cite{PS,ACPZ},
rigidity term \cite{P0} etc. These terms contain  powers of
$\partial/m_\gamma$. For thin strings $m_\gamma$ is large
 and these corrections can be neglected in the action (\ref{ng}).
However, as we explained in the Introduction
for the ANO vortex in the semiclassical regime
$\partial^2/m^2_\gamma\sim T/m^2_\gamma\sim1/g^2\gg1$ (see
(\ref{tint})). Hence, higher derivative corrections blow up in (\ref{ng})
and the string approximation is not  acceptable. In particularly,
there is no world sheet conformal invariance.

The  QED coupling is large at the AD point  so there is no hope to
to use quasiclassical analysis to derive the sting theory for
ANO flux tubes from QED. Therefore, we take another route.
We have to construct this string theory imposing the world sheet
conformal invariance. The latter requirement follows from (\ref{smallt}).
To see this note, that if there were no world sheet conformal invariance,
the string tension would get renormalized \cite{P0} and become
of order of  the string theory UV cutoff, which is $v^2$. This is in
 a contradiction with (\ref{smallt}).

To maintain the world sheet conformal invariance is quite a problem for
the string moving in the space with   a non-critical dimensionality.
The Liouville coordinate does not decouple and a string should be
considered as moving in the 5 dimensional space \cite{P81}. Even this
``high price'' appeared to be not enough. For $d\ge 1$ the string turns
out to be unstable.

 The way out  was found  by
Polyakov who conjectured that the fifth coordinate should be curved
 \cite{P1,P}.
The bosonic part of the string action looks like \cite{P1,P}
\beq
\label{str}
S_{\rm str}=\ T_0\int d^2\sigma \left[ a(y)(\partial_i x^{\mu})^2
+ (\partial_i y)^2 + \Phi(y)R_2 + \cdots \right].
\eeq
Here $y$ is the Liouville coordinate, $R_2$ is the world sheet curvature
and $\Phi$ is the dilaton depending on the fifth coordinate only.
Dots stands for other possible background fields.
Of course, the metric in the 4-dimensional slice of this space should be
flat.

The function $a(y)$ gives the running string tension
\beq
\label{aten}
T\ =\ T_0 \;a(y),
\eeq
where $T_0$ is the classical string tension (\ref{ct}) which we consider as 
UV data at the UV scale $v$.
It is  subject to a renormalization (\ref{aten}) within the string theory
(\ref{str}).

The condition of the world sheet conformal invariance means the
vanishing of $\beta$-functions for the 2d theory (\ref{str}). The latter
conditions coincide with  equations of motion for the 5d effective
``gravity'' \cite{CMPF}
\beq
\label{5g}
S_{gr}=\frac{1}{2\kappa}\int d^5 x\sqrt{g}\exp{(-2\Phi)}\left[ R+
2(D_M \Phi)^2 + 4V(0)+\cdots\right] .
\eeq
Here $M=1,...5$, $R$ is 5d curvature, $\Phi$ is the dilaton 
and $V(0)$ sands for the 
``cosmological constant'' which is the value of the scalar potential
 at zero. Dots in (\ref{5g}) represents other background fields
of the string theory (\ref{str}) to be included in (\ref{5g}).

Generically (\ref{5g}) is relatively useless because we don't know 
terms associated with these additional background fields. However,
 we can look at it as at an  effective low energy
theory at scales below the string scale $\sqrt{T}$. Then all string
states can be integrated out and we are left with an effective
theory of ``gravity''
for a few light fields. In this setup string is considered
as moving in a slow varying classical gravitational background.
 After finding a solution of gravitational 
equations of motion we have to check that the  curvature of
the 5d space is much smaller than the string scale $T$.

 Following \cite{M,P} we 
include in (\ref{5g}) the  $U(1)$ R-R 5-form $F$ besides the 
5d metric $g_{MN}$ and the dilaton $\Phi$.
With these fields taken into account  Einstein
equations of motion take  the form
$$
R_{MN}+ 2D_M D_N \Phi + T\exp{(2\Phi)}\left[F_{MKLPQ}F_{N}^{KLPQ}
\right.
$$
\beq
\label{mn}
 \left. -\frac{1}{10}g_{MN}F_{SKLPQ}F^{SKLPQ}\right]=0,
\eeq
while the equation balancing the central charge is
\beq
\label{cch}
R-4(D_M \Phi)^2 +4D_M D^M \Phi + 4V(0)+\pi T(10-d)=0,
\eeq
where $D_M$ is the covariant derivative. The last term in
(\ref{cch}) comes from the anomaly \cite{P81}.
It is nonzero for the non-critical dimension $d=5$.
Since we started from \ntwo QED we assume a target space supersymmetry
for the string theory (\ref{str}), thus the critical dimension would
correspond to $d=10$..
The equation for the $U(1)$ 5-form reads
\beq
\label{feq}
D^M F_{MKLPQ}=0
\eeq

The metric in our 5d ``gravity'' has a special form  determined
by the single function $a(y)$ (see (\ref{str}))
\beq
\label{met}
 ds^2=  \;a(y)(dx_{\mu})^2 +(dy)^2\;.
\eeq

 With this ansatz for the metric eqs.
(\ref{mn})-(\ref{feq}) were studied in \cite{P}. 
The solution for (\ref{feq})
corresponds to a constant 5-form field strength
\beq
\label{fsol}
F_{MKLPQ}=\frac{f}{\sqrt{g}}\epsilon_{MKLPQ}=\frac{f}{a^2}\epsilon_{MKLPQ},
\eeq
where $f$ is a dimensionless constant. Within the large $N$ approach
of \cite{M} $f\sim N$, so it is natural to assume that  $f\sim 1$ in
the case at hand.
 Substituting this result back into eqs.
(\ref{mn}), (\ref{cch}) we get
\beq
\label{munu}
-\frac{1}{2}\frac{a''}{a}-\frac{1}{2}\frac{a'^2}{a^2}
+\frac{a'}{a}\Phi '=Tf^2\exp{(2\Phi)}
\eeq
for $\mu\nu$ components of (\ref{mn}) and
\beq
\label{55}
-2\frac{a''}{a}+\frac{a'^2}{a^2}
+2\Phi ''=Tf^2\exp{(2\Phi)}
\eeq
for the  $55$ component.
The equation for zero central charge becomes \footnote{In fact, it is a
certain  linear combination of the equation (\ref{cch}) and eqs.
(\ref{munu}), (\ref{55}).}
\beq
\label{an}
-\frac{1}{2} \Phi'' + \Phi '^2 
-\frac{a'}{a}\Phi'-\frac{5}{4}Tf^2\exp{(2\Phi)}
=\, V(0)+\frac{5\pi}{4}T
\eeq
Here prime stands for the derivative with respect to $y$.

These equations admit \cite{P} the solution with AdS metric
\beq
\label{ads}
a(y)=\exp{ (2\frac{y}{r_0})}
\eeq
and constant dilaton
\beq
\label{condil}
\Phi=\Phi_0.
\eeq
Here $r_0$ is the radius of the AdS space. We can trust our ``gravity''
solution if this radius is large enough in string units \cite{M},
\beq
\label{r0con}
r_0\gg 1/\sqrt{T}.
\eeq
Substituting (\ref{ads}) into eqs. (\ref{munu})-(\ref{an}) we find
\beq
\label{gs}
f^2\exp{(2\Phi_0)}=-\frac{4}{T\, r_0^2 }
\eeq
and
\beq
\label{ft}
r_0^2=\frac{20}{4V(0)+ 5\pi T}.
\eeq

From (\ref{ft}) we see that in order to fulfill the condition (\ref{r0con})
we need some cancelation between the ``cosmological constant'' term $V(0)$
 and the contribution due to the anomaly in  (\ref{ft}). Unfortunately, we
don't know the scalar potential for our 5d ``gravity'' and cannot
check (\ref{r0con}) on the gravity side.
 \footnote{In principle,
certain information about scalar potential can be extracted
 by the dimensional 
reduction from 10d supergravity \cite{FGPW}. We do not follow this 
approach in this paper because our 5d ``gravity'' is an effective
low energy theory which has nothing to do with real 10d gravity.}
We will come back to this issue below
and  use information on the field theory side to
show that (\ref{r0con}) is fulfilled.

Now let us discuss the meaning of the Liouville coordinate $y$ in the 
4d field theory.  It was suspected for a long time 
that this coordinate has
a meaning of RG scale from the point of view of 4d field theory. In 
\cite{M} this interpretation was formulated explicitly. To be more
specific, it is convenient to introduce a new coordinate $r$ instead of
$y$ as
\beq
\label{r}
r=r_0\exp{\frac{y}{r_0}}.
\eeq
In terms of this coordinate the AdS metric looks like
\beq
\label{adsmet}
 ds^2= \;\frac{r^2}{r_0^2}(dx^{\mu})^2 +
r_0^2\frac{(dr)^2}{r^2}\;.
\eeq
Now following \cite{M} we introduce the coordinate
\beq
\label{u}
u=T\,r
\eeq
which has dimension of energy and identify it with the energy scale
of the RG flow, $u=\mu$ (see \cite{PP} for the discussion
of the normalization in this identification).

In particularly, large $u$, $u\ga u_0 $ (here $u_0=Tr_0$) 
 corresponds
to UV region in the field theory, while moving to small $u$ 
towards the throat of the AdS space is associated with the RG
flow to the IR. As the mass of the photon ($\sim v$) serves
as a UV cutoff for our  effective string theory of ANO string,
it is natural to identify
\beq
\label{u0}
u_0=v.
\eeq
This identification shows immediately that the AdS radius is
large in string units,
\beq
\label{lr0}
r_0=\frac{u_0}{T}= \frac{v}{T}\gg\frac{1}{\sqrt{T}},
\eeq
where we use the condition of smallness of the 
string tension (\ref{smallt}).
 This result shows that  we can trust the ``gravity'' solution.

It was shown in \cite{M,GKP,W} that AdS metric in the 5d gravity
  corresponds to a conformal invariant field theory in 4d. 
To understand this note, that the solution (\ref{ads}), (\ref{condil})
has constant dilaton and zero values of other fields
which we do not include in the effective 5d ``gravity''(\ref{5g}).
These fields play the role of ``coupling constants'' in the 4d
field theory \cite{GKP,W}. As soon as coupling constants
do not run with the RG energy scale $u$ we are dealing with CFT.
We obtained our  string theory under condition (\ref{smallt})
as an effective theory at the AD point on the Higgs branch.
Therefore the conformal theory in question is the $CFT_{AD}^{H}$
which we discussed in the previous subsection.

Now let us see what does the ``gravity'' solution with AdS metric
(\ref{adsmet}) gives for the renormalized string tension  
(\ref{aten}). Clearly, 4d conformal invariance means that $T=0$.
This was explicitly shown in \cite{M2} by the calculation of the  Wilson
loop in the $AdS_5$ background. This result is qualitatively clear
from eq. (\ref{aten}) with function $a(y)$ given by (\ref{ads}).
The string goes all the way down to the throat of the AdS space at 
$u=0$ ($y=-\infty$) producing the zero result for the tension.

To get a non-zero string tension we have to prevent the 
string from penetrating into the throat. This can be done by the conformal
symmetry breaking at some scale $u_{CB}$ associated with the  kink 
(domain wall) solution
of 5d ``gravity'' separating AdS region at large $u$  from
a different AdS regime 
 in the IR at small $u$ (cf. \cite{FGPW,GPPZ0,GPPZ}). We consider
 this breaking in the next section.

Note, in conclusion of this section that the result $T=0$ which follows
from the AdS solution of the 5d ``gravity'' coincides with our field
theory expectations (\ref{dec}) showing selfconsistency of our approach.

\section{Deformation of the AdS metric}
\setcounter{equation}{0}

In general, the 4d  CFT can be driven away from criticality by a relevant
scalar operator $O$ with conformal dimension $\Delta\le 4$ by
adding the term
\beq
\label{odef}
\int d^4 x\, \sigma_0\, O(x)
\eeq
to the action. Here $\sigma_0$ is a ``coupling constant''. From the 
5d ``gravity'' point of view this constant 
becomes a scalar field $\sigma(r)$
with the boundary value $\sigma=\sigma_0$ at the UV boundary
$r=r_0$ \cite{GKP,W}. Near the boundary at large $r$ it behaves as
\beq
\label{assbeh}
\sigma=\sigma_0 \left( \frac{r_0}{r}\right)^{4-\Delta}.
\eeq

In the theory at hand we associate the breaking of the conformal invariance
with moving slightly away from AD point. Then the monopole mass
becomes non-zero, see (\ref{mm}). Thus, the relevant deformation in 
question is the monopole mass term with the conformal dimension 
$\Delta=2$. On the  ``gravity'' side we have to include the scalar field
$\sigma$ in our 5d ``gravity'' (\ref{5g}) with the boundary condition
\beq
\label{sig0}
\sigma_0=\frac{m_m^2}{u_0^2},
\eeq
which is determined by the small monopole mass $m_m$ at the scale $u=u_0$.

 Einstein equations of motion modify as
$$
-\frac{1}{2}\frac{a''}{a}-\frac{1}{2}\frac{a'^2}{a^2}
+\frac{a'}{a}\Phi '=Tf^2\exp{(2\Phi)},
$$
$$
-2\frac{a''}{a}+\frac{a'^2}{a^2}
+2\Phi '' +2\sigma'^2=Tf^2\exp{(2\Phi)},
$$
\beq
\label{scE}
-\frac{1}{2} \Phi'' + \Phi '^2 
-\frac{a'}{a}\Phi'-\frac{5}{4}Tf^2\exp{(2\Phi)}
= V(\sigma)+\frac{5\pi}{4}T.
\eeq
Also we have additional equation for $\sigma$
\beq
\label{sigeq}
- \sigma '' + 2\Phi ' \sigma '
-2\frac{a'}{a}\sigma ' +\frac{\partial V(\sigma)}{\partial \sigma}=0.
\eeq
Here $V(\sigma)$ is the potential for the scalar $\sigma$ which we
unfortunately don't know.

 Below we develop a perturbation theory
near the AdS metric at large $r$ to make an estimate of 
the scale $u_{CB}$ at which
the kink solution takes over and destroys the AdS metric.
We assume that the scalar potential has the following expansion
\beq
\label{pot}
V(\sigma)= V(0) + \frac{1}{2}m_{\sigma}^2\sigma^2 +
\frac{\lambda}{r_0^2}\sigma^4 + \cdots ,
\eeq
where $\lambda$ is dimensionless constant and $m_{\sigma}$ is the mass
of the scalar $\sigma$. Eq. (\ref{sigeq}) gives for this mass
\beq
\label{msi}
m_{\sigma}^2=-\frac{4}{r_0^2}
\eeq
in accordance with the general result \cite{GKP,W}
\beq
\label{del}
\Delta = 2+ \sqrt{4+ m_{\sigma}^2 r_0^2}
\eeq
relating the conformal dimension of operator $O$ in 4d field theory and
the mass of the corresponding scalar field in 5d gravity \footnote{
Note that the value of mass in (\ref{msi}) is at the border of
stability, see \cite{BrF}.}.

The solution of equations of motion  (\ref{scE}) to the second order
in the perturbation around the AdS metric looks like
$$
a(y)=e^{2\frac{y}{r_0}}\left[\; 1+ 
(\frac{\lambda}{8}-\frac{1}{25})\sigma_0^4
e^{-8\frac{y}{r_0}} + \cdots \right],
$$
$$
\Phi=\Phi_0-\frac{\sigma_0^2}{5}e^{-4\frac{y}{r_0}}+
\frac{\sigma_0^4}{25} e^{-8\frac{y}{r_0}}+\cdots,
$$
\beq
\label{gsol}
\sigma= \sigma_0 e^{-2\frac{y}{r_0}}+ (\frac{\lambda}{4}-\frac{1}{5})
\sigma_0^3  e^{-6\frac{y}{r_0}} + \cdots.
\eeq
Here the metric , the dilaton and the scalar $\sigma$ expressed in
terms of the boundary value $\sigma_0$ which is determined by the
small monopole mass via the boundary condition (\ref{sig0}). The 
 dilaton expectation value $\Phi_0$
is constrained by eq. (\ref{gs}).
To find higher terms in this expansion (or exact solution for the kink)
 we need to know the form of  the potential (\ref{pot}).
 
Note, that generally speaking the equation (\ref{sigeq}) admits
two solutions with behavior $r_0^2/r^2$ and $log(r_0^2/r^2)r_0^2/r^2$
at large $r$. One can show, however, that the second solution does
not lead to a consistent solution of  other equations of motion in
(\ref{scE}).

For generic couplings $\lambda$( $\lambda \sim 1$) corrections in
(\ref{gsol}) becomes of order of one and destroy the AdS metric at
$y_{CB}$ determined by
\beq
\exp{(2\frac{y_{CB}}{r_0})}\sim \sigma_0.
\eeq
In terms of the RG scale variable $u$ the above equation reads
\beq
\label{ucb}
\frac{u_{CB}^2}{u^2_0}\sim \sigma_0.
\eeq
Substituting here the boundary value of $\sigma_0$ (\ref{sig0}) we get
the scale of the conformal symmetry breaking
\beq
\label{ucb2}
\frac{u_{CB}^2}{u^2_0}\sim \frac{m_m^2}{v^2}.
\eeq

Now let us make an estimate of the value of the renormalized string
tension at the scale  $u_{CB}$. Eq. (\ref{aten}) gives
\beq
\label{adsten}
T= T_0\frac{u^2}{u_0^2},
\eeq
where we use $a(u)=u^2/u_0^2$  for the AdS metric.
Assuming that the kink stops the string from penetrating into
the throat of the AdS space and determines the scale $u_{CB}$ to be
substituted into eq. (\ref{adsten}) we get 
\beq
\label{rt}
T\sim \frac{m_m^2}{\log vL},
\eeq
where we use (\ref{ucb2}) and the expression (\ref{ct}) for the bare
string tension at the UV scale $v$. Recall that the length of the 
string $L$ appears here because we do not have infinitely long 
flux tubes on the Higgs branch where quarks are strictly massless,
 see subsection 2.2.

The estimate (\ref{rt}) is our final result for the string tension
near the AD point. To prove it rigorously (and to work out the coefficient
in (\ref{rt})) one has to find the exact solution for the kink and to
calculate the Wilson loop in the kink background (cf. \cite{M2,GPPZ0,GPPZ}).
This is left for a future work.

Note, that the result (\ref{rt}) satisfy conditions (\ref{smallt}) and
(\ref{dec}) showing the consistency of the 
  5d ``gravity'' description with field theory expectations.
In particular, (\ref{rt}) gives zero tension at the AD point. As we
already explained, the string becomes tensionless, however still remains
to be a stringy object because its transverse size $\sim v^{-1}$ is
finite. This is an interesting example of the non-trivial
conformal  theory $CFT_{AD}^{H}$ containing massless quarks
and tensionless ANO strings with massless monopoles attached to 
 ends of  these  strings.

Now let us estimate the value of the string coupling constant and show
that it is small, $g_s\ll 1$. First note, that as we already mentioned
in section 3  the radius of the  AdS space is large in string units.
 Substituting  (\ref{rt}) into (\ref{lr0}) we get
\beq
\label{larr0}
r_0\sim \frac{v}{m_m^2}.
\eeq
where we drop the logarithm factor.
Now eq. (\ref{gs}) gives for the closed
string coupling constant ($g_s^2=\exp{2\Phi_0}$)
\beq
\label{grgs}
g_s^2 f^2\,\sim \,\frac{m_m^2}{v^2}\ll 1.
\eeq
As we already mention we
 assume that  the field strength of the RR 5-form
is of order one, $f\sim 1$. Then (\ref{grgs}) gives $g_s\ll 1$.

We can also estimate the string coupling from the field theory
side. The open string coupling constant
measures the probability that the string
is broken by the monopole-anti-monopole pair production. This
probability is of order of \cite{NSY}
\beq
\label{stdec}
g_s\sim \exp{(-c\frac{m_m^2}{T})},
\eeq
where the positive constant $c$ can be, in principle calculated.
Taking into account the logarithm factor in (\ref{rt})
we get
\beq
\label{ftgs}
g_s \sim \left(\frac{m_m}{v}\right)^{\gamma}\ll 1,
\eeq
where $\gamma > 0$. Here we use that the typical length of the string
is of order of $L\sim 1/\sqrt{T}$ for small hadron spins. The reason why
the string coupling constant turns out to be small is that monopoles,
although light with respect to photon, appear to be heavy with
respect to the string scale $\sqrt{T}$ because of the logarithm factor
in (\ref{rt}). Therefore, the monopole-anti-monopole production is
suppressed.

Although we don't know $\gamma$ in the field theory expression
(\ref{ftgs}) as well as we don't know
$f^2$ on the ``gravity'' side, the two 
estimates (\ref{ftgs}) and (\ref{grgs}) are consistent with each other and 
show that our string theory is in the weak coupling.

\section{Discussion}

In this paper we considered ANO flux tubes on the Higgs branch near the
AD point in \ntwo QCD. The effective low energy 
 QED describing the theory near the root of the Higgs branch
becomes strongly coupled when we approach
 the AD point. Thus, the semi-classical analysis is no longer valid.
We present  arguments based on 
 the consideration of instanton induced higher derivative
corrections on the Higgs branch 
that the ANO string tension is small, much smaller than the scale
determined by the quark condensate, see (\ref{smallt}), (\ref{dec}).
This condition ensures that the ANO flux tube is   long and thin
and can be described by the non-critical string theory with the 
world sheet conformal invariance. This leads us to the string 
moving in 5d space with curved fifth (Liouville) coordinate \cite{P1,P}.
At the AD point the 4d conformal invariance on the field theory side
corresponds to AdS background metric of the 5d space. We also
considered the breaking of the conformal invariance by moving
slightly away from the AD point. From 5d ``gravity'' equations of
motion we found that the 
renormalized ANO string tension is determined by the small mass
of monopoles, see (\ref{rt}).

The main lesson to learn is that once we make the confining matter
light the ANO string becomes light too and does not 
``want'' to live in four dimensions.
It goes into a five dimensional space. This happens
already at the hadron scale.

Note, that we do not use the logic of large $N$ in this paper \cite{M}.
The radius of the AdS space remains unfixed within the 5d ``gravity''
approach, see (\ref{ft}). Instead, we use the field theory arguments
to show that it is large in string units, see (\ref{larr0}). Thus the
 ``gravity'' solution can be trusted.

At first sight it  seems surprising that 5d ``gravity'' gives reasonable
results. It seems to ``know'' almost nothing about the field theory
at the AD point under consideration. Still the results we obtained
from 5d ``gravity'' are consistent with the field theory expectations.
 This means that the 5d ``gravity''
description is quite universal. The only information from the field
theory side that we use formulating the string theory (\ref{str}) and
its effective  5d ``gravity'' description is the 
existence of two scales $v$ and $m_m$, subject to  the condition
$m_m\ll v$ as well as 
 relations (\ref{smallt}), (\ref{dec}) for the string tension.
On the 5d ``gravity'' side we identified the UV scale $v$ with the  radius
of the AdS space $u_0$ ( see (\ref{u0})) and the small monopole mass
$m_m$ with the boundary value of the  5d scalar $\sigma$, see 
(\ref{sig0}). Then  Einstein equations give  the estimate
(\ref{ucb2}) which tells us that the scale of the conformal symmetry
breaking is given by $m_m$, which is quite obvious conclusion
from  the field theory point of view.
 Then the result (\ref{rt}) comes from the expression for 
the running string tension (\ref{adsten}) in the AdS background
\cite{P1,P}.

Of course, to find the exact solution for the kink 
(domain wall) we need to use
much more information about our field theory near AD point. First,
we have to impose \ntwo supersymmetry. On the 5d ``gravity'' side
this means imposing ${\cal N} =4$ supersymmetry for the AdS background
which corresponds to $CFT_{AD}^{H}$ at large $u$ (cf. \cite{FGPW}).
The kink should be  1/2-BPS solution preserving the \ntwo supersymmetry.
 It should satisfy  first order
differential equations. This kink  separates the UV AdS region
associated with  $CFT_{AD}^{H}$ from the different 
 AdS regime in the IR. The latter should
correspond to the free theory of massless quark moduli on the Higgs branch
which emerges at scales well below the string scale $\sqrt{T}$.
We also have to impose global $SU(2)_{R}$
symmetry which becomes a gauge symmetry in 5d ``gravity''. Hopefully
this information would be enough to fix the scalar potential
in 5d ``supergravity'' and find the kink solution.

Another important open problem is related to the
 presence of massless quarks on the Higgs branch. The number of
massless quark moduli specify precisely the $CFT_{AD}^{H}$ we are
dealing with. In particular, it determines
fractional conformal dimensions of various operators.
The presence of massless quark moduli
 should be taken into account in the 5d ``gravity'' description.
In ref. \cite{AFM} fractional  dimensions were related
to the spectrum of Laplacian operator in the geometry determined by
3-branes moving near 7-branes singularities. It would be interesting
to find a 5d analog of this description.

\subsection*{Acknowledgments}

The author is grateful to Alexandr Gorsky,
Gennady Danilov, Andrei Marshakov, Erich Poppitz,  Mikhail Shifman,
 Arkady Vainshtein and, in particular, to  Alexei Morozov
 for helpful discussions. The author also would like
to thank the Theoretical Physics Institute at the University of Minnesota
for hospitality and support. This work is also supported by Russian
Foundation for Basic Research under grant No. 99-02-16576 and by the 
US Civilian Research and Development Foundation under grant 
No. RP1-2108.


\begin{thebibliography}{99}

\bibitem{SW1}N.~Seiberg and E.~Witten, hep-th/9407087;
Nucl.Phys.B {\bf426} (1994) 19.

\bibitem{SW2}N. Seiberg and E. Witten, hep-th/9408099;
Nucl.Phys.B {\bf431} (1994) 484.

\bibitem{MH}S. Mandelstam, Phys.Rep. {\bf23C} (1976) 145;\\
G.'t Hooft, in Proceed.of the Europ.Phys.Soc. 1975, ed.by
A.Zichichi (Editrice Compositori, Bologna, 1976), p.1225.

\bibitem{ANO}A.A. Abrikosov, Sov.Phys.JETP, {\bf32} (1957)
1442;\\
H.B. Nielsen and P. Olesen, Nucl.Phys. {\bf B61} (1973) 45.

\bibitem{HSZ} A. Hanany, M. Strassler, and A.~Zaffaroni,
hep-th/9707244; Nucl.Phys.B {\bf513} (1998) 87.

\bibitem{Y} A. Yung, hep-th/9906243; Nucl.Phys. B{\bf562} (1999)
191.

\bibitem{YR} A. Yung, hep-th/0005088; Proceedings of
the XXXIV PNPI Winter School, Repino, St. Petersburg, 2000.

\bibitem{BF} A. Bilal and F. Ferrari, hep-th/9706145;
Nucl.Phys.B {\bf516} (1998) 175.

\bibitem{AD}P.C. Argyres and M.R. Douglas, hep-th/9505062;
Nucl.Phys.B {\bf448} (1995) 93.

\bibitem{APSW}P.C. Argyres, M.R. Plesser, N.~Seiberg, and
E.~Witten, hep-th/9511154; Nucl.Phys.B {\bf461} (1996) 71.

\bibitem{hori} T. Eguchi, K. Hori, K. Ito, and S. Yang,
hep-th/9603002;  Nucl.Phys.B {\bf471} (1996) 430.

\bibitem{P1} A. Polyakov, hep-th/9711002;  Nucl.Phys.
Proc. Suppl.  {\bf68} (1998) 1. 

\bibitem{P} A. Polyakov, hep-th/9809057; Int. J. Mod. Phys.
 A {\bf14} (1999) 645. 

\bibitem{M} J. Maldacena, hep-th/9711200; Adv. Theor. Math. Phys.
{\bf2} (1998) 231.

\bibitem{GKP} S. S.Gubser, I. R. Klebanov and A. M. Polyakov,
 hep-th/9802109;  Phys. Lett. B  {\bf428} (1998) 105. 

\bibitem{W} E. Witten, hep-th/9802150; Adv. Theor. Math. Phys.
{\bf2} (1998) 253.

\bibitem{VY} A. Vainshtein and A. Yung, {\em Type I superconductivity
upon monopole condensation in Seiberg-Witten theory}, hep-th/0012250;

\bibitem{APS} P. Argyres, M. Plesser and N. Seiberg, hep-th/9603042;
Nucl. Phys. B {\bf471} (1996) 159.

\bibitem{Y2} A. Yung, hep-th/9705181; Nucl. Phys. B {\bf512} (1998) 79.
 
\bibitem{Y3} A. Yung, hep-th/9605096;  Nucl. Phys. B {\bf485} (1997) 38. 

\bibitem{R}A. Penin, V. Rubakov, P.~Tinyakov and S.~Troitsky,
hep-th/9609257; Phys.Lett B {\bf389} (1996) 13.

\bibitem{ARH} A. Achucarro, M. de Roo and L.~Huiszoon,
Phys.Lett. B{\bf424} (1998) 288.

\bibitem{IS} K. Intrilligator and N. Seiberg, hep-th/9509066;
Nucl. Phys. BC (Proc. Suppl.) {\bf45} (1996) 1.

\bibitem{GVY} A. Gorsky, A. Vainshtein and A. Yung, hep-th/0004087;
 Nucl. Phys. B {\bf584} (2000) 197.

\bibitem{O} P. Orland, hep-th/9404140; Nucl.Phys. B{\bf428}
(1994) 221;\\ M. Sato and S. Yahikozawa, hep-th/9406208;
Nucl.Phys.B {\bf436} (1995) 100.


\bibitem{BS}M. Baker and R. Steinke, hep-ph/9905375;
Phys. Lett. B {\bf474} (2000) 67

\bibitem{PS}J. Polchinski and A. Strominger, Phys.Rev.Lett.
{\bf67} (1991) 1681.

\bibitem{ACPZ}E. Akhmedov, M. Chernodub, M.~Polikarpov, and
M.~Zubkov, Phys.Rev.D {\bf53} (1996) 2087.

\bibitem{P0}A. Polyakov, Nucl.Phys.B {\bf268} (1986) 406.

\bibitem{CMPF} C. Callan, E. Martinec, M. Perry and D. Friedan,
Nucl.Phys.B {\bf262} (1985) 593.

\bibitem{P81}A. Polyakov, Phys. Lett. B {\bf103} (1981) 207.

\bibitem{FGPW} D. Z. Freedman, S. S. Gubser, K. Pilch and
N. P. Warner, {\em Renormalization group flows from holography-
supersymmetry and a c-theorem},  hep-th/9904017.

\bibitem{PP} A. Peet and J. Polchinski, hep-th/9809022;
Phys. Rev. D {\bf59} (1999) 065011.

\bibitem{M2} J. Maldacena, hep-th/9803002; Phys. Rev. Lett.
{\bf80} (1998) 4859.

\bibitem{GPPZ0} L. Girardello, M. Petrini, M. Poratti and A. Zaffaroni,
hep-th/9903026; JHEP {\bf9905} (1999) 026.


\bibitem{GPPZ} L. Girardello, M. Petrini, M. Poratti and A. Zaffaroni,
hep-th/9909047; Nucl.Phys.B {\bf569} (2000) 451.

\bibitem{BrF} P. Breitenlohner and D. Z. Freedman, Phys. Lett. B
{\bf115} (1982) 197; Ann. Phys. {\bf144} (1982) 197.

\bibitem{NSY} I. Novikov, M. Shifman and A. Yung, to appear.

\bibitem{AFM} O. Aharony, A. Fayyazuddin and J. Maldacena,
hep-th/9806159; JHEP {\bf9807} (1998) 013.





















\end{thebibliography}
\end{document}